\documentclass{cimento}
\usepackage{graphicx}

\title{GRB\,030329: Three Years of Radio Afterglow Monitoring}

\author{A.J.~van~der~Horst\from{ins:1}\ETC, 
A.~Kamble\from{ins:2}, 
R.A.M.J.~Wijers\from{ins:1}, 
L.~Resmi\from{ins:2}\from{ins:3}, \\
D.~Bhattacharya\from{ins:2}, 
E.~Rol\from{ins:4}, 
R.~Strom\from{ins:5}\from{ins:1}, 
C.~Kouveliotou\from{ins:6}, \\
T.~Oosterloo\from{ins:5} \atque 
C.H.~Ishwara-Chandra\from{ins:7}}

\instlist{\inst{ins:1} Astronomical Institute, University of Amsterdam, Kruislaan 403, 1098 SJ Amsterdam, The~Netherlands
\inst{ins:2} Raman Research Institute, Bangalore 560080, India
\inst{ins:3} Joint Astronomy Programme, Indian Institute of Science, Bangalore 560012, India
\inst{ins:4} Department of Physics and Astronomy, University of Leicester, University Road, Leicester LE2 7RH, UK
\inst{ins:5} ASTRON, P.O. Box 2, 7990 AA Dwingeloo, The~Netherlands
\inst{ins:6} NASA/MSFC, NSSTC, VP62, 320 Sparkman Drive, Huntsville, AL 35805, USA
\inst{ins:7} National Center for Radio Astrophysics,  Post Bag 3, Ganeshkind, Pune 411007, India}

\PACSes{\PACSit{98.70.Rz}{gamma-ray sources; gamma-ray bursts}
\PACSit{95.85.Bh}{Radio, microwave ($>1$ mm)}}

\begin{document}

\maketitle

\begin{abstract}
Radio observations of gamma-ray burst (GRB) afterglows are essential for our understanding of the physics of relativistic blast waves, as they enable us to follow the evolution of GRB explosions much longer than the afterglows in any other wave band. We have performed a three-year monitoring campaign of GRB\,030329 with the Westerbork Synthesis Radio Telescopes (WSRT) and the Giant Metrewave Radio Telescope (GMRT). Our observations, combined with observations at other wavelengths, have allowed us to determine the GRB blast wave physical parameters, such as the total burst energy and the ambient medium density, as well as investigate the jet nature of the relativistic outflow. Further, by modeling the late-time radio light curve of GRB\,030329, we predict that the Low Frequency Array (LOFAR, 30-240~MHz) will be able to observe afterglows of similar GRBs, and constrain the physics of the blast wave during its non-relativistic phase.
\end{abstract}

\section{GRB\,030329 Radio Afterglow}
GRB\,030329 displayed one of the brightest afterglows ever, enabling the study of its evolution for a long time and in detail over a broad range of frequencies, from X-ray to centimetre wavelengths. The afterglow was still visible in radio waves 1100 days after the burst trigger. The optical afterglow was visible for only a couple of weeks until it was obscured by the emerging supernova associated with the GRB; the X-ray afterglow was detected for eight months. Thus using our radio observations we can uniquely determine the physical parameters of the blast wave.
The broadband afterglow can be modeled by the standard relativistic blast wave model assuming that the jet has two components: one component with a small opening angle that explains the early-time optical and X-ray light curves, and a wider component that carries the bulk of the energy and produces the later-time light curves. Because the peak and self-absorption frequency of the broadband spectrum of the second jet are situated in the centimetre regime, this component is naturally best studied at these wave bands.

\section{WSRT \& GMRT Results}
The afterglow of GRB\,030329 was observed with the WSRT and the GMRT from 325~MHz to 8.4~GHz \cite{ref:1}. We have modeled the light curves we obtained together with earlier reported fluxes from WSRT \cite{ref:2}, GMRT \cite{ref:3}, and VLA \& ATCA \cite{ref:4}\cite{ref:5}. The afterglow was clearly detected at all frequencies except for 325~MHz, where we could only obtain upper limits. GRB\,030329 is the first afterglow to be detected at frequencies below 1~GHz: at 840~MHz with WSRT and even as low as 610~MHz with GMRT. 

The light curves show the peak of the broadband synchrotron spectrum moving to lower frequencies in time. After 80 days the observed light curves decrease less steeply than expected, which can be explained by a transition into the non-relativistic phase of the blast wave \cite{ref:2}\cite{ref:3}\cite{ref:5}. It was suggested in \cite{ref:2} that this late-time behaviour could also be explained by a third jet-component with an even wider opening angle than the first two. However, the latter model is excluded by the observations below 1~GHz, which leaves the model with the non-relativistic phase after 80 days as the preferred model for the late-time behaviour of the blast wave \cite{ref:1}.

\section{LOFAR}
The Low Frequency Array will be a major new multi-element, interferometric, imaging telescope designed for the 30-240~MHz frequency range. LOFAR will use an array of simple omni-directional antennas. The electronic signals from the antennas are digitised, transported to a central digital processor, and combined in software to emulate a conventional antenna. LOFAR will have unprecedented sensitivity and resolution at metre wavelengths. This will give the GRB community the opportunity to study bright afterglows on even longer timescales than with observations at centimetre wavelengths. 

We have extrapolated the modeling results of the radio afterglow of GRB\,030329 to the LOFAR observing range \cite{ref:1}. The predicted light curves show that GRB\,030329 will be observable in the high band of LOFAR (120-240~MHz), but not in the low band (30-80~MHz). We also calculated light curves for GRB\,030329 if it were situated at a redshift of 1 instead of 0.16. The resulting fainter afterglow can also be detected in the high band, although with longer integration times (on the order of a day instead of an hour).




\begin{thebibliography}{5}
\bibitem{ref:1} \BY{Van~der~Horst~A.J., Kamble~A., Wijers~R.A.M.J. \etal}
   2006, in preparation
\bibitem{ref:2} \BY{Van~der~Horst~A.J., Rol~E., Wijers~R.A.M.J. \etal}
   \IN{ApJ}{634}{2005}{1166}
\bibitem{ref:3} \BY{Resmi~L., Ishwara-Chandra~C.H., Castro-Tirado~A.J. \etal}
  \IN{A\&A}{440}{2005}{477}
\bibitem{ref:4} \BY{Berger~E., Kulkarni~S.R., Pooley~G. \etal}
  \IN{Nature}{426}{2003}{154}
\bibitem{ref:5} \BY{Frail~D.A., Soderberg~A.M., Kulkarni~S.R. \etal}
  \IN{ApJ}{619}{2005}{994}
\end{thebibliography}
\end{document}